\documentclass{article} \usepackage{amsmath, amsthm, amssymb, graphicx, url}
  \usepackage{setspace}
  
\title{Modified Papoulis-Gerchberg algorithm for sparse signal recovery}
\author{M.H. Kayvanrad (mohammad@nus.edu.sg), D. Zonoobi, A.A. Kassim}
\begin{document}
\date{}
\maketitle

\doublespacing


\begin{abstract}
 Motivated by the well-known Papoulis-Gerchberg algorithm, an iterative thresholding algorithm for recovery of sparse signals from few observations is proposed. The sequence of iterates turns out to be similar to that of the thresholded Landweber iterations, although not the same. The performance of the proposed algorithm is experimentally evaluated and compared to other state-of-the-art methods.
\end{abstract}

\section{Introduction and Problem Definition} \label{sec: intro}

In many real world problems, instead of the complete signal, we have some observations of the signal of interest from which we want to reconstruct the original signal. In the simplest case, which is fortunately applicable to many practical situations, the observation process can be approximated as a linear operator:
\begin{equation}
o=Kf
\end{equation}
where $f$ is the original signal of interest, $o$ is the observed features, i.e. samples, and $K$ is the (linear) observation operator. We are interested in the case where the number of observations is much fewer than the length of the original signal. That is, $K_{m \times n}$ has much fewer rows than columns, i.e. $m \ll n$.  Furthermore, in practice the observation process is not exact and typically we can only obtain a distorted version of the observed features. This distortion is usually modeled by an additive error term. So the observation process can be modeled as
\begin{equation} \label{eq: lin inv prob}
g=o+e=Kf+e
\end{equation}
in which $e$ represents the (additive) observation error, e.g. noise.

The objective is to find the original signal, $f$, from the set of available observations, $g$, while $K$ is also known. That is, to solve the linear inverse problem \eqref{eq: lin inv prob}. In order to do so one might minimize the discrepancy $\Delta(f)=\|Kf-g\|^{2}$. However, except for the very special case that the operator $K$ has a trivial null space (see \cite{daubechies: iterative thresholding} for details), the minimizer is not unique. In order to address this problem, i.e. to regularize the inverse problem, one might suppose a priori assumptions and impose different constraints on the solution, which is usually taken into account by adding a penalization term to the discrepancy. In this letter we are especially interested in the case where we have a sparsity constraint on the solution. For the sake of a more precise explanation, suppose there exists an orthonormal basis $(\psi_{i})$ in which the signal of interest can be expanded with the expansion coefficients $\theta_{i}=<f,\psi_{i}>$, a large number of which are zero or negligibly small. In order to make use of this constraint for solving the inverse problem \eqref{eq: lin inv prob}, define the $l^{p}$-norm $\| \theta \|_{p}=(\sum_{i}|\theta_{i}|^{p})^{1/p}$. One might then find the minimizer of the following functional, as the solution of \eqref{eq: lin inv prob} with the aforementioned sparsity constraint:
\begin{equation} \label{eq: regularized inv prob}
\Phi_{\mu}(f)=\|Kf-g\|^{2}+\mu \| \theta \|_{p}
\end{equation}
where $0<p<2$ and $\mu$ is the regularization parameter that can be chosen based on application.

This is a well-known problem and different approaches to solving it have been proposed. Here we will not go through the details of the problem, which have been widely studied by other authors. The reader is referred to \cite{daubechies: iterative thresholding} for a comprehensive discussion of the problem. For the sake of consistency, we follow the same mathematical notations as those used in \cite{daubechies: iterative thresholding}.
Furthermore, it is worth mentioning that beside the $lp$-norm introduced above, some authors have as well used the total variation (TV) norm as the constraint. See, for example, \cite{candes: robust uncertainty principles}.


There are several methods of recovery available in the literature, among of which iterative thresholding algorithms are an important class. Iterative thresholding algorithms are, more or less, based on a thresholded version of the Landweber iterations (see, for example, \cite{book: intro to math theo of inv probs}). I.e. the sequence of iterates has the general form
\begin{equation} \label{eq: threshlded landweber iterations}
f^{n}=S_{\gamma}(f^{n-1}+K^{*}(g-Kf^{n-1}))
\end{equation}
where $K^{*}$ denotes the conjugate of $K$, and $S_{\gamma}$ is a thresholding operator.
In \cite{daubechies: iterative thresholding} the authors prove the convergence of the above iterative algorithm to the (unique) minimizer of \eqref{eq: regularized inv prob}, when $S_{\gamma}$ is the soft thresholding operator (see the definition of soft thresholding in section \ref{sec: description of algorithm}).
Noticeable effort has been put into accelerating the original algorithm. In \cite{paper: domain decomposition methods} the authors propose a method for accelerating thresholded Landweber iterations, which is based on alternating subspace corrections. Other methods for this purpose are introduced in \cite{paper: acceleration of projected gradient method}, and \cite{paper: a new twist}. Although use of soft thresholding is more common, some authors have, as well, used \emph{hard} thresholding to address the above inverse problem. See \cite{paper: iterated hard shrinkage for minimization problems}, \cite{paper: iterative thresholding for cs} or \cite{paper: iterative thresholding algorithms} as examples.

\section{Description of the proposed algorithm} \label{sec: description of algorithm}

In order to explain the underlying idea of the proposed method, let us begin with the following problem. In \cite{papoulis: siganal} Papoulis introduces \emph{an iteration method} for reconstruction of a band-limited signal from a known segment. Suppose $f(t)$ is a signal of which we only know a small segment, $g(t)=p_{\tau}(t)f(t)$, where $
p_{\tau}(t)= \left\{ \begin{array}{ll}
1 & \textrm{$|t|\leq \tau$}\\
0 & \textrm{otherwise}
\end{array} \right.
$. Also suppose $F(\omega)$ is the Fourier transform of $f(t)$ and $F(\omega)=0$ for $|\omega|>\sigma$ (bandlimitedness). The objective is to reconstruct $f(t)$ from $g(t)$.

In order to solve this problem, we begin with $G(\omega)$, the Fourier transform of $g(t)$, and form $H_{1}(\omega)=G(\omega)p_{\sigma}(\omega)$, i.e. truncate $G(\omega)$ for $|\omega|>\sigma$. In other words, we change $g$ so that it satisfies the constraint on the original signal (bandlimitedness in this case). $h_{1}(t)$, the inverse transform of $H_{1}(\omega)$, is then used to form $f_{1}(t)=g(t)+h_{1}(t)-h_{1}p_{\tau}(t)$, which recovers the known segment of $f$. $f_{1}(t)$ is supposed to be a better estimate of the desired signal, $f(t)$, than $g(t)$. This estimate can be further improved by repeating the above procedure in an iterative manner. That is, in the $n$th iteration, we form the function:
\begin{equation}
H_{n}(\omega)=F_{n-1}(\omega)p_{\sigma}(\omega)
\end{equation}
compute its inverse transform, $h_{n}(t)$, and recover the known segment of the original signal
\begin{equation} \label{eq: papoulis method-recovery of original segment}
f_{n}(t)=g(t)+h_{n}(t)-h_{n}(t)p_{\tau}(t)
\end{equation}
It can be proved that $F_{n}(\omega)$ tends to $F(\omega)$ as $n \rightarrow \infty$ \cite{papoulis: siganal}.

In brief, in each iteration, we change the latest estimate of the desired signal, i.e. the output of the previous iteration, so that it satisfies the constraint (bandlimitedness in this case). Since this process might affect the entire signal, including the known segment, the known segment is then recovered before further progress.

This problem is obviously different from our original problem stated in \eqref{eq: regularized inv prob}, because, firstly, it concentrates on the special case of recovering a \emph{continuous} signal from \emph{a known segment} and, secondly, the constraint on the signal is \emph{bandlimitedness} while the constraint of \eqref{eq: regularized inv prob} is \emph{sparsity}. Nevertheless, we will implement the above idea to solve our own problem as explained below.

Based on the above algorithm, our \emph{iterative} algorithm involves two main operations in each iteration, namely, an operation to maintain the constraint followed by an operation to recover the original observations. Since we are interested in problems with sparsity constraint, a thresholding operation can maintain this constraint for us, i.e.
\begin{equation}
h^{n}=S_{\gamma}(f^{n-1})
\end{equation}
where $f^{n-1}$ is the latest estimate of the original signal, obtained in the previous iteration, and $S_{\gamma}$ is the \emph{soft} thresholding operator, defined as
\begin{equation} \label{eq: thresholding operator}
S_{\gamma}(g)= \sum_{i}s_{\gamma}(<g,\psi_{i}>)\psi_{i}
\end{equation}
where $s_{\gamma}(x)= \left \{ \begin{array}{ll}
x+\frac{\gamma}{2} & \textrm{$x \leq -\frac{\gamma}{2}$} \\
0 & \textrm{$|x|<\frac{\gamma}{2}$} \\
x-\frac{\gamma}{2} & \textrm{$x \geq \frac{\gamma}{2}$}
\end{array} \right. $

Analogous to \eqref{eq: papoulis method-recovery of original segment}, the original observations are then recovered by
\begin{equation}
f^{n}=h^{n}+K^{*}(g-Kh^{n})
\end{equation}

The sequence of iterates can, thus, be expressed in the following form:
\begin{equation} \label{eq: sequence of iterates}
f^{n}=K^{*}(g+S_{\gamma}(f^{n-1})-KS_{\gamma}(f^{n-1}))
\end{equation}
with $f^{0}=K^{*}g$.

Although \eqref{eq: sequence of iterates} is not exactly a sequence of Landweber iterations, it can still be viewed as a modified version of the thresholded Landweber iterations. Note, especially, the analogy between \eqref{eq: sequence of iterates} and \eqref{eq: threshlded landweber iterations}.

In this letter we only introduce the algorithm and experimentally evaluate its performance, compared to similar state-of-the-art algorithms. A detailed discussion of the convergence of the iterative algorithm and its relation to the thresholded Landweber iterations is beyond the scope of the current letter and will be postponed to future publications. The motivation behind the proposed algorithm was briefly discussed, though.

\section{Experiments}

In all the experiments described below, the thresholding operator is applied to stationary wavelet transform (SWT) \cite{swt ref} coefficients, obtained using DB1 (Haar) mother function for 1 level of decomposition. All thresholds are obtained using the well-known Birge-Massart strategy \cite{birge-massart strategy ref}. The iterative algorithm continues until a convergence criterion, e.g. $\|x^{k+1}-x^{k}\|/\|x^{k}\|<\delta$, is met. For the sake of comparison, the results are compared with those obtained by $L_1$ norm minimization \cite{candes: intro to cs} and total-variation (TV) norm minimization \cite{candes: robust uncertainty principles}, which are two well-known state-of-the-art methods of sparse signal recovery.

Due to space constraints, the results of the experiments are included very concisely. More comprehensive results can be found at \url{http://mkayvan.googlepages.com/sparsesignalrecovery}.

\subsection{Recovery of 1D signals} \label{exp: 1d}

\begin{figure}
\includegraphics[height=30mm]{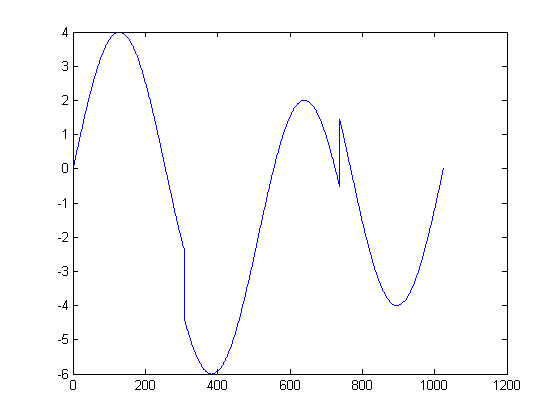}
\includegraphics[height=30mm]{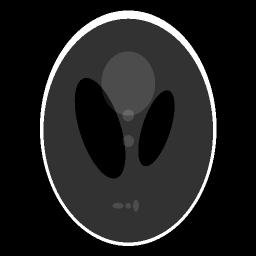}
\caption{Left: HeaviSine test signal; right: Shepp-Logan Phantom}
\label{fig: heavisine phantom}
\end{figure}

First, we consider the ideal case of sampling with no distortion, i.e. we assume $e=0$ in \eqref{eq: lin inv prob}.
The HeaviSine test signal (figure \ref{fig: heavisine phantom}), from the well-known Donoho-Johnstone \cite{donoho-johnstone collection ref} collection of synthetic test signals, is reconstructed from different numbers, $M$, of randomly selected samples.
Table \ref{tbl: mse nonoise 1d heav} shows the the mean squared error (MSE) between the reconstructed and the original signal for reconstruction by \eqref{eq: sequence of iterates} as well as by $L1$ and TV norm minimization. As it is obvious from results, reconstruction by \eqref{eq: sequence of iterates} outperforms the two other methods in almost all cases.


\begin{table}
\centering
\begin{tabular}{|c| c c c|}
\hline
$\#$Sample & Rec. by \eqref{eq: sequence of iterates} & $L1$ norm & T.V. norm\\
\hline \hline
$M=70$ & $0.0339$ & $0.13$ & $0.0547$ \\                                
$M=100$ & $0.055871$ & $0.096$ & $0.024$ \\
$M=150$ & $6.51e-03$ & $0.038$ & $0.0131$ \\
$M=200$ & $4.65e-03$ & $0.03$ & $0.01$ \\
\hline
\end{tabular}
\caption{MSE between the reconstructed and the original signal for reconstruction by \eqref{eq: sequence of iterates} as well as by $L1$ and TV norm minimization, from $M$ observed samples. The original HeaviSine test signal constitutes $N=1024$ samples.}
\label{tbl: mse nonoise 1d heav}
\end{table}

\subsection{Recovery of 2D Signals}

Here we consider the classical problem of reconstruction of images from highly incomplete frequency domain observations, which has received considerable attention, because of its important applications in medical imaging, especially in MRI.
In particular, acquiring MR images involves acquisition of 2-D Fourier domain data of the image. Due to the physics of the imaging device, this process is usually carried out by taking 1-dimensional slices from the 2-dimensional Fourier domain data of the image. This process is often too time-consuming, though, so for a rapid MR imaging it is desirable to take only a subset of these slices, e.g. a reduced number of Fourier domain samples taken over radial lines.

Radial sampling is a common way of sampling over the 2-dimensional Fourier domain, and several authors have addressed this problem, especially as an important application of compressive sampling. For the sake of comparison of the proposed method with other state-of-the-art reconstruction methods, here we will also address the same problem. The reconstruction problem might, however, seem a bit different here, since sampling is confined to radial lines, instead of being random. Nevertheless, what we are essentially doing is reconstruction of the signal from its projections onto a lower dimensional subspace, which is exactly what was being done in the previous cases.

The test image used is the Shepp-Logan phantom (figure \ref{fig: heavisine phantom}) of size  $256 \times 256$. The pixels in this image take values between 0 and 1, and the image has a nonzero gradient at 2184 pixels. The setup of our experiments is the same as that of \cite{candes: robust uncertainty principles} which has been as well adopted by several other authors as a framework to evaluate the performance of their methods, including \cite{paper: gradient pursuits}, \cite{paper: recursive spatially adaptive filtering}, \cite{paper: reweighted l1 minimization}, and \cite{paper: stagewise weak gradient pursuits part 1}.

\begin{figure}
\includegraphics[width=29mm]{phantom.jpg}
\includegraphics[width=29mm]{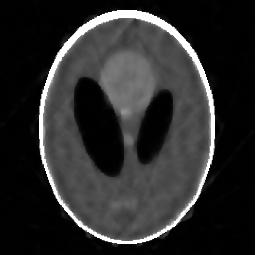}
\includegraphics[width=29mm]{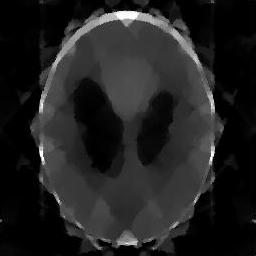}
\includegraphics[width=29mm]{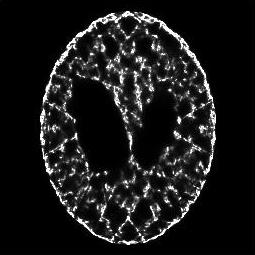}
\caption{Reconstruction of the $256 \times 256$ Shepp-Logan phantom from samples taken along $K=9$ radial lines in the Fourier domain. From left to right: (a) Original $256 \times 256$ Shepp-Logan phantom (b) Reconstruction by \eqref{eq: sequence of iterates} (c) Reconstruction by TV norm minimization (d) Reconstruction by $l1$ norm minimization.}
\label{fig: rec phantom 9 lines}
\end{figure}

Table \ref{tbl: phantom noiseless reconstruction} shows the PSNR (peak signal-to-noise ratio) values of the reconstructed images from samples taken along $k=9$, 11, 15, and 21 radial lines in the 2-dimensional discrete Fourier Transform (DFT) domain, compared to those obtained by $L1$ and TV norm minimization. The reconstructed images by \eqref{eq: sequence of iterates} as well as by minimizing $l1$ and TV norms, from samples taken over 9 radial lines, are shown in Figure \ref{fig: rec phantom 9 lines}.

\begin{table}
\centering
\begin{tabular}{|c| c c c|}
\hline
$\#$Radial Lines & Rec. by \eqref{eq: sequence of iterates} & T.V. norm & $L1$ norm \\
\hline \hline
$K=9$ &  $24.9746$ &          $14.36$     & $11.8$\\
$K=11$ & $29.2307$ &   $21$        & $13.45$  \\
$K=15$ & $39.3145$ &          $21.66$     & $15.33$ \\
$K=21$ & $199.7471$ &  $113.2541$  & $27.1$ \\
\hline
\end{tabular}
\caption{Reconstruction of the $256 \times 256$ Shepp-Logan Phantom from samples along $K$ radial lines in the 2D-DFT domain. Reconstruction quality is measured in terms of PSNR of the reconstructed image.}
\label{tbl: phantom noiseless reconstruction}
\end{table}

In the next experiment we consider the more realistic case of distorted observations. Particularly, samples are taken from noisy versions of the Shepp-Logan Phantom test image, affected by additive white Gaussian noise (AWGN). The results are shown in table \ref{tbl: phantom noisy 31 lines}. Note that especially when the noise is significant, the reconstructed image enjoys considerably less error than the noisy one from which we got our samples.

\begin{table}
\centering
\begin{tabular}{|c| c c c c|}
\hline
dB & Noisy image & Rec. by \eqref{eq: sequence of iterates} & $L1$ norm & T.V. norm\\
\hline \hline
$20$ & $20.0663$ & 26.2296 & $18.65$ & $18.01$ \\
$30$ & $30.0176$ & 34.8228 & $22.94$ & $25.89$ \\
$40$ & $40.0028$ & 43.7792 & $25.43$ & $32.34$ \\
$50$ & $50.0185$ & 53.4432 & $26.15.$ & $48.8$ \\
\hline
\end{tabular}
\caption{Reconstruction of the $256 \times 256$ Shepp-Logan Phantom from noisy samples affected by AWGN. The first column shows the PSNR of the AWGN; the second column shows the PSNR of the noise-affected image, from which the samples are taken, with respect to the original image. The corresponding values of PSNR, for reconstruction by \eqref{eq: sequence of iterates}, $L1$ norm minimization, and TV norm minimization are shown in the third, fourth, and fifth columns, respectively.}
\label{tbl: phantom noisy 31 lines}
\end{table}

\section{Conclusion}
Motivated by the Papoulis-Gerchberg algorithm, a method for recovery of sparse signals from very limited numbers of observations was proposed. Iterative thresholding algorithms have been widely used to address this problem. Our algorithm also takes advantage of thresholding to maintain the sparsity constraint in each iteration. The signal is then reconstructed by iteratively going through a constraint-maintaining operation followed by recovery of the known features. The performance of the method was experimentally evaluated and compared to other state-of-the-art methods.


\end{document}